\documentclass[12pt]{article}

\usepackage[english]{babel}

\usepackage{latexsym}
\usepackage{amssymb}
\usepackage{epsfig}

\usepackage{amsmath}
\usepackage{graphics}
\usepackage{graphicx}
\usepackage{subfigure}

\begin{document}

\title{Anomalous Transmission in Waveguides with Correlated Disorder in Surface Profiles}

\author{ F.~M.~Izrailev${}^{1}$ and N.~M.~Makarov${}^{2}$ \\
{\it ${}^{1}$ Instituto de F\'{\i}sica, Universidad Aut\'{o}noma de Puebla} \\
{\it Puebla, Pue., 72570, Mexico}\\
{\it ${}^{2}$ Instituto de Ciencias, Universidad Aut\'{o}noma de Puebla} \\
{\it Priv. 17 Norte No. 3417, Col. San Miguel Hueyotlipan}\\
{\it Puebla 72050, M\'{e}xico}}

\maketitle



\section{Introduction}

In recent years an increasing attention is paid to the so-called {\it correlated disorder} in low-dimensional disordered systems. The interest to this subject is mainly due to two reasons. First, it was found that specific correlations in a disordered potential can result in a quite unexpected anomalous properties of scattering. Second, it was shown that such correlations can be relatively easily constructed experimentally, at least, in the one-dimensional Anderson model and in Kronig-Penney models of various types. Therefore, it seems to be feasible to fabricate random structures with desired scattering properties, in particular, when one needs to suppress or enhance the localization in given frequency windows for scattering electrons or electromagnetic waves. In addition, it was understood that in many real systems the correlated disorder is an intrinsic property of underlying structures. One of the most important examples is a DNA chain for which strong correlations in the potential have shown to manifest themselves in an anomalous conductance. Thus, the subject of correlated disorder is important both from the theoretical viewpoint, and for various applications in physics.

The key point of the theory of correlated disorder is that the localization length for eigenstates in one-dimensional models absorbs the main effect of correlations in disordered potentials. This fact was known since the earliest analytical studies of transport in continuous random potentials, however, until recently the main interest was in delta-correlated potentials, or in potentials with a Gaussian-type of correlations. On the other hand, it was shown \cite{IK99,IKU01,ML98,IM05} that the most interesting effect is related to specific long-range correlations that can be fabricated in practice. In particular, it was demonstrated that in the standard one-dimensional Anderson model one can observe effective {\it mobility edges} in the energy spectrum, when the pair correlator computed along the disorder is of specific form. The important feature of the mobility edge $\omega_c$ in the frequency spectra of traveling waves, is that it separates the region of strongly localized states from that of very extended states. This property is important for the construction of potentials with selective transmission or reflection. From the experimental viewpoint, many of the obtained results may have a strong impact for the creation of a new class of electron nanodevices, optic fibers, acoustic and electromagnetic waveguides with selective transport properties.

In spite of an asymptotic character of theoretical results obtained for infinitely large samples and weak disorder, the analytical predictions were found to work relatively good for a strong disorder as well. The first experimental study \cite{KIKS00,KIKSU02,KIK08} of both the suppression and enhancement of localization due to correlated disorder, was performed on single-mode electromagnetic waveguides. It was shown that in the case of statistically correlated point-like surface scatterers one can create the controlled frequency windows of the enhanced transmission, or windows with a very strong reflection. The important point is that in spite of many experimental imperfections, such as very strong absorbtion or small number of scatterers with large amplitudes, the selective transport was clearly observed in accordance with the theory.

For single-mode waveguides the problem of the surface scattering can be reduced to that for one-dimensional disordered models. For this
reason the methods and results obtained for the latter case can
be directly applied for the waveguides \cite{IM01,IM02,IM03,IM04a,IM04b}. The situation is principally different for many-mode waveguides.
The problem of wave propagation through such systems with corrugated
surfaces has a long history and till now remains a hot topic
in the literature. This problem naturally arises in the analysis
of spectral and transport properties of optics fibers, acoustic
and radio waveguides, remote sensing, shallow water waves,
multilayered systems and photonic lattices, etc.
\cite{BFb79,DB86,O91,FG92,S95,ON95}. Similar problems emerge in quantum physics, when describing the propagation of
quasi-particles in thin metal films and semiconductor
nanostructures, such as nanowires and strips, superlattices and
quantum-well-systems \cite{C69,B73,LGPb88,MKb04,BH90,L92,FM94,D95,Do98}.  Recently, new theoretical results \cite{IM09,IMT09} have been obtained for one-dimensional disordered models describing photonic crystals, bi-layered metamaterials and electronic superlattices. The experiments performed on microwave guiding systems with intentionally randomized model parameters \cite{LIMKS09}, have confirmed the predictions of the theory.

As is well established, the scattering from corrugated surfaces
results in the diffusive transport
\cite{CE69,TJM86,TA88,FC89,KMY91,MMY95,ZB95,SXW95,MS94}, as well as in the effects of strong electron/wave localization \cite{MM84,BCHMM85,MY89,TF92,MT98,BN96,GTSN98,LFL98,SFYM98}.
Correspondingly, the eigenstates of periodic systems with
corrugated surfaces turn out to have a chaotic structure
\cite{LKRK96}. Recent numerical studies of quasi-one-dimensional
{\it surface-disordered} systems \cite{GTSN98,LFL98,SFYM98} have revealed a principal difference from those known in the standard models with
{\it bulk} random potentials \cite{FM94}. Specifically, it was
found that transport properties of quasi-one-dimensional waveguides with rough
surfaces essentially depend on many characteristic lengths. In comparison, for the bulk scattering all transport characteristics depend on one parameter only, which is the ratio of the localization length to the size of a sample (the so-called {\it single-paramter scaling}).

The situation in many-mode waveguides in the presence of long-range correlations was found to be quite tricky
\cite{IM03,IM03a,IM04}. It was shown that
the long-range correlations, on the one hand, give rise to a
suppression of the interaction between different propagating
waveguide modes. On the other hand, the same correlations can
provide a perfect transparency of each independent channel,
similar to what happens in the one-dimensional geometry. The number of
independent transparent modes is governed by the correlation
length and can be equal to the total number of propagating
modes. Therefore, the transmission through such waveguides can be
significantly enhanced in comparison with the case of uncorrelated
surface roughness.

It should be stressed that the main results in the theory of surface
scattering were obtained for random surfaces with fast-decaying
correlations along the structures. Therefore, it is of great
importance to explore the role of specific long-range correlations
in surface profiles, using the results found for one-dimensional
systems with correlated disorder. Apart from the theoretical
interest, this problem can be studied experimentally since the existing
experimental technics allow for the construction of systems with
sophisticated surfaces \cite{WOD95,KS98}.

We would like to note that in order to focus on the role of long-range correlations in surface profiles, in this review we do not discuss the so-called {\it square-gradient} mechanism of scattering. As was recently shown, this mechanism emerges due to a quite specific dependence of the back scattering length on the second derivative of scattering profiles. The theoretical aspects of the square-gradient scattering and its possible experimental implications can be found in Refs.\cite{SGS}.

\section{Surface-corrugated waveguide}

\begin{figure}[h]
{\includegraphics[width=\columnwidth]{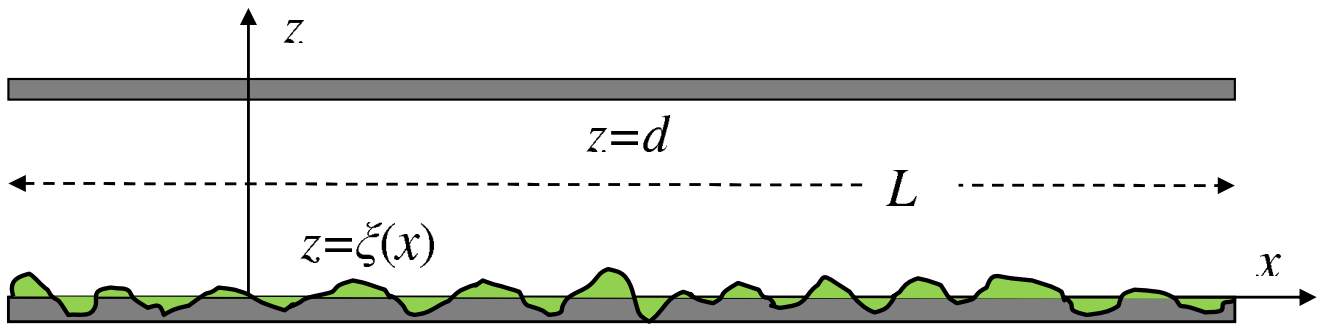}}
\caption{(Color online) Plane waveguide with lower
corrugated edge.}
\label{Fig1-SCW}
\end{figure}

As a physically plausible and commonly used model to study the multiple surface scattering, we consider an open plane waveguide of length $L$ and average width $d$ with perfectly conducting lateral walls. It is naturally to imply the waveguide length be much greater than its width, $d\ll L$. Such a system is called quasi-one-dimensional one. The $x$-axis is stretched along the structure and the $z$-axis is directed in transverse direction. The lower boundary of the waveguide is supposed to have rough (corrugated) profile $z=\xi(x)$, slightly deviated from its flat average $z=0$. The upper boundary is taken to be flat, $z=d$, (see Figure~\ref{Fig1-SCW}). Thus, the surface-corrugated guiding system occupies the area defined by the relations,

\begin{equation}\label{SD-region}
-L/2<x<L/2, \qquad\qquad \xi(x)\leqslant z\leqslant d.
\end{equation}

The random function $\xi(x)$ describing the surface roughness is assumed to be statistically homogeneous with the following characteristics,

\begin{equation}\label{SD-Ksi}
\langle\xi(x)\rangle=0,\qquad\langle\xi^2(x)\rangle=\sigma^2,\qquad
\langle\xi(x)\xi(x')\rangle=\sigma^2{\cal W}(x-x').
\end{equation}
The angular brackets $\langle\ldots\rangle$ stand for ensemble averaging over the disorder, i.e. over different realizations of the random surface profile $\xi(x)$, or for the spatial average over the  coordinate $x$ of any prescribed realization. Both types of the average are assumed to be equivalent due to ergodicity. The variance of $\xi(x)$ is denoted by $\sigma^2$ and, consequently, $\sigma$ is the root-mean-square roughness height. The binary (two-point) correlator ${\cal W}(x)$ is normalized to its maximal value, ${\cal W}(0)=1$, and is assumed to decrease with increasing $|x|$ on characteristic scale termed the \emph{correlation length}.

In what follows we consider weak surface scattering for which the corrugations are small, $\sigma\ll d$. This limitation is common in the surface scattering theories based on appropriate perturbative approaches \cite{BFb79}. As is known, for weak scattering all transport properties are entirely determined by the {\it roughness power spectrum} $W(k_x)$,

\begin{subequations}\label{FT-W}
\begin{eqnarray}
{\cal W}(x)&=&\int_{-\infty}^{\infty}\frac{dk_x}{2\pi}
\exp\left(ik_xx\right)\,W(k_x),\\[6pt]
W(k_x)&=&\int_{-\infty}^{\infty}dx\exp(-ik_xx)\,{\cal W}(x).
\end{eqnarray}
\end{subequations}
Since the pair correlator ${\cal W}(x)$ is a real and even function of the coordinate $x$, its Fourier transform $W(k_x)$ is even, real and non-negative function of the lengthwise wave number $k_x$. Note that the condition ${\cal W}(0)=1$ is equivalent to the following normalization for $W(k_x)$,

\begin{equation}\label{W-norm}
\int_{-\infty}^{\infty}\frac{dk_x}{2\pi}\,W(k_x)=1.
\end{equation}

Since in the $x$-direction the system is open, at $x=\pm L/2$ we assume the radiative boundary conditions. In the transverse $z$-direction the zero Dirichlet boundary conditions are applied to both lateral walls, $z=\xi(x)$ and $z=d$. Thus, the analysis of the surface scattering in our model is reduced to the study of the following two-dimensional boundary-value problem,

\begin{subequations}\label{2D-BVP}
\begin{eqnarray}
\left(\frac{\partial^2}{\partial x^2}+\frac{\partial^2}{\partial
z^2}+k^{2}\right)\Psi(x,z)=0,\\[6pt]
\Psi(x,z=\xi(x))=0,\qquad \Psi(x,z=d)=0.
\end{eqnarray}
\end{subequations}
Here $k=\omega/c$ is the total wave number for electromagnetic wave of frequency $\omega$ and TE polarization, propagating through a waveguide with perfectly conducting walls. Note that in contrast with the bulk-scattering, here the wave equation does not contain any scattering potential since the source of the scattering is the roughness of boundaries.

In ideal waveguide with flat walls, $\xi(x)=0$, the solution of the problem \eqref{2D-BVP} has the canonical form of normal waveguide modes,

\begin{equation}\label{2D-NWM}
\Psi_{n,\pm}(x,z)=\frac{1}{\sqrt{\pi d}}\,
\sin\left(\frac{\pi nz}{d}\right)\,\exp(\pm ik_nx).
\end{equation}
Here the integer $n=1,2,3,\ldots$ enumerates the normal modes \eqref{2D-NWM} with the transverse wave number $k_z=\pi n/d$. The lengthwise wave number $k_x=\pm k_n$ for the $n$th mode is,

\begin{equation}\label{kn}
k_n=\sqrt{k^2-(\pi n/d)^2}.
\end{equation}
Evidently, the transport properties depend only on
normal modes that can propagate along the waveguide, i.e. have real value of $k_n$. As follows from Eq.~\eqref{kn}, such \emph{propagating modes} have indices $n\leqslant N_d$, and their total number $N_d$ is equal to the integer part $[...]$ of the ratio $kd/\pi$,

\begin{equation}\label{Nd}
N_d=[kd/\pi].
\end{equation}
The waveguide modes with indices $n>N_d$ have purely imaginary wave numbers $k_n$. These \emph{evanescent modes} decay exponentially fast on the scale of the order of wavelength. As one can see, the unperturbed (flat) waveguide is equivalent to the set of $N_d$ one-dimensional non-interacting \emph{conducting channels} occupied by the corresponding propagating modes.

\section{Single-mode structure}

Keeping in mind the relevance of wave scattering to the Anderson localization, let us, first, consider a single-mode waveguide. In this case the mode parameter $kd/\pi$ is restricted by the relation $1<kd/\pi<2$, and the number of conducting channels equals one, $N_d=1$. The transmission through such waveguide depends on the lengthwise wave number $k_1$,

\begin{equation}\label{k1}
k_1=\sqrt{k^2-(\pi /d)^2}.
\end{equation}
All other waveguide modes with $n\geqslant2$ are evanescent and do not contribute to the transport properties. From the single-mode
condition it follows that the wave number $k_1$ is
confined within the interval,

\begin{equation}\label{SM-k1}
0<k_1d/\pi<\sqrt{3}.
\end{equation}
Note that the weak surface-scattering condition $\sigma\ll
d$, leads to the inequality $k_1\sigma\ll 1$.

As was shown in Refs.~\cite{MY89}, the transport problem \eqref{2D-BVP} for the surface-disordered single-mode
waveguide is equivalent to a one-dimensional disordered model,

\begin{equation}\label{1DSchr-eq}
\left[\frac{d^{2}}{dx^{2}}+k^{2}-V(x)\right]\psi(x)=0
\end{equation}
where $k_1$ stands in place of $k$. In such a description, the effective potential $V(x)$ has the form,

\begin{equation}\label{SD-SM-V}
V(x)=\frac{2}{\pi}\,\left(\frac{\pi}{d}\right)^3 \,\xi(x),
\end{equation}
which is entirely determined by the rough surface profile $\xi(x)$.

As one can see, the surface scattering in one-mode waveguides is equivalent to the bulk scattering emerging in one-dimensional disordered systems. The latter problem can be solved with the use of well-developed methods,
such as the perturbative diagrammatic technique of
Berezinski \cite{B73}, the invariant imbedding method
\cite{MY89,BWb75,Kb86} or the two-scale approach \cite{MT98,M99}. All these methods allow one to take adequately into account the effects of coherent multiple scattering from the corrugated surface giving rise to the Anderson localization.

The main theoretical result is that the average transmittance
$\langle T\rangle$ as well as all its moments $\langle T^s\rangle$, are described by the universal function,

\begin{eqnarray}\label{<1D-Ts>}
\langle T^s(L/L_{loc})\rangle&=&\frac{1}{2\sqrt{\pi}}
\left(\frac{L}{2L_{loc}}\right)^{-3/2}
\exp{\left(-\frac{L}{2L_{loc}}\right)}\nonumber\\[6pt]
&\times&\int_0^\infty\frac{zdz}{\cosh^{2s-1}z}
\exp\left(-z^2\frac{L_{loc}}{2L}\right) \int_0^z
dy\cosh^{2(s-1)}y\,,
\nonumber\\[6pt] &&\ s=0,\pm1,\pm2,\ldots\,.
\end{eqnarray}
This function depends solely on the ratio $L/L_{loc}$ between the waveguide length $L$ and the \emph{localization length} $L_{loc}$ (see, for example, Ref.~\cite{LGPb88}).

In one-dimensional disordered systems a wave can be scattered either forward or backward. However, it was shown that the transport properties are determined exclusively by the backscattering while the forward scattering has no influence. Therefore, the quantity $L_{loc}$ is, in fact, the {\it backscattering length} emerging in an infinite one-dimensional structure. It is important that the inverse value $L_{loc}^{-1}$ can be associated with the Lyapunov exponent appearing in various transfer matrix approaches \cite{LGPb88}. In the latter description the Lyapunov exponent gives the average rate of decrease of the wave function $\langle\psi(x)\rangle$ away from the center of its localization.

It should be stressed that the dependence of the transport properties on the ratio $L/L_{loc}$ manifests a principal concept of \emph{one-parameter scaling} that constitutes the phenomenon of one-dimensional Anderson localization. The non-trivial point of this concept is that in order to describe the transport properties of {\it finite} samples of size $L$, it is sufficient to know how the wave function is localized in an {\it infinite} sample with the same disorder.

From Eq.~\eqref{<1D-Ts>} one can derive relatively easy the expressions for low moments of the transmittance $T$.
Specifically, at $s=1$ one obtains the average transmittance $\langle T(L/L_{loc})\rangle$,

\begin{eqnarray}\label{1D-avT}
\langle T(L/L_{loc})\rangle&=&\frac{1}{2\sqrt{\pi}}
\left(\frac{L}{2L_{loc}}\right)^{-3/2}
\exp{\left(-\frac{L}{2L_{loc}}\right)}\nonumber\\[6pt]
&&\times\int_0^\infty\frac{z^2dz}{\cosh z}
\exp\left(-z^2\frac{L_{loc}}{2L}\right).
\end{eqnarray}
The second moment, $s=2$, is important to get the variance of the transmittance. It can be shown that for $L_{loc}\lesssim L$, the variance is of the order of squared average transmittance itself. This means that for strong localization the transmittance is not self-averaged quantity. Hence, by changing the length $L$ of the waveguide, or the disorder itself, one should expect very large fluctuations of the transmittance. Such fluctuations are known as the \emph{mesoscopic fluctuations} that are characteristic of strong interference effects on macroscopic scale.

In order to properly characterize the transport properties of
one-dimensional structures for any degree of localization (ratio $L/L_{loc}$), one should refer to the self-average logarithm of the transmittance,

\begin{equation}\label{1D-avLog}
\langle\ln T(L/L_{loc})\rangle=-2L/L_{loc}.
\end{equation}
This result is consistent with an exponential decrease of the transmittance averaged over the so-called representative (non-resonant) realizations of the random disorder \cite{LGPb88},

\begin{equation}\label{1D-repavT}
\langle T(L/L_{loc})\rangle_{rep}=\exp(-2L/L_{loc}).
\end{equation}

Note that Eq.~\eqref{1D-avLog} quite often is used as the definition of the localization length $L_{loc}$ itself. It is highly non-trivial that by exploring the transmission properties of finite samples, one can extract the localization length that is defined for infinite samples. This fact is again the manifestation of the one-parameter scaling.

In accordance with the scaling concept, in the one-dimensional geometry there are only two characteristic regimes corresponding to the ballistic and localized transport.

\textbf{(i)} The \emph{ballistic transport} occurs if the localization length $L_{loc}$ is much larger than the system length $L$. In this case one-dimensional structure is practically fully transparent since its average transmittance is close to one,

\begin{equation}\label{1D-Tbal}
\langle T(L/L_{loc})\rangle\approx1-2L/L_{loc}
\qquad\mbox{for}\qquad L_{loc}\gg L.
\end{equation}
This asymptotic expression results from both expressions, \eqref{1D-avT} and \eqref{1D-repavT}.

\textbf{(ii)} Otherwise, the disordered structures exhibit the \emph{localized transport}, when the localization length $L_{loc}$ is smaller than the sample length $L$. In this case the average transmittance \eqref{1D-avT} is exponentially small,

\begin{eqnarray}\label{1D-Tloc}
\langle T(L/L_{loc})\rangle&\approx&\frac{\pi^3}{16\sqrt{\pi}}
\left(L/2L_{loc}\right)^{-3/2}
\exp\left(-L/2L_{loc}\right)\nonumber\\[6pt]
&&\mbox{for}\qquad L_{loc}\ll L.
\end{eqnarray}
This means that in the localization regime the disordered one-mode waveguides perfectly (with an exponential accuracy) reflect the incoming waves.

According to the above expression the transmission exponentially decreases on the scale $L \approx 2L_{loc}$, with an additional power prefactor. In contrast, the transmittance \eqref{1D-repavT} has an exponential dependence with much faster decrease on the scale $L\approx L_{loc}/2$. This fact can be explained as follows. The main contribution to the asymptotic form \eqref{1D-Tloc} for the average transmittance \eqref{1D-avT} is given by resonant realizations of the random potential $V(x)$. For these realizations the transmittance is almost equal to one, however, they have an exponentially small probability. On the other hand, for representative realizations (most probable, but non-resonant) the transmittance is described by formula \eqref{1D-repavT}. This effect is peculiar to the mesoscopic nature of Anderson localization.

The expressions \eqref{<1D-Ts>} -- \eqref{1D-Tloc} are universal and applicable for any one-dimensional system with a weak static disorder. As one can see, in order to describe transport properties of finite structure, one needs to know the localization length $L_{loc}$. According to different approaches \cite{B73,LGPb88,BWb75,Kb86,M99}, the inverse localization length for any kind of weak disorder is determined by the $2k$-harmonic in the randomness power spectrum $S(k_x)$ of the scattering potential $V(x)$,

\begin{subequations}\label{1D-Lloc}
\begin{eqnarray}
L_{loc}^{-1}(k)&=&S(2k)/8k^2\,;\\[6pt]
&&\langle V(x)V(x')\rangle=C(x-x'),\\[6pt]
&&S(k_x)=\int_{-\infty}^{\infty}dx\exp(-ik_xx)\,C(x).
\end{eqnarray}
\end{subequations}
For elastic backward scattering the wave vector $\vec{k}$ conserves its value $k$, changing the sign, $|\Delta\vec{k}|=2k$. Accordingly, Eq.~\eqref{1D-Lloc} reflects the fact that the localization length is defined by the backscattering length only.

The expression \eqref{1D-Lloc} indicates that the global properties of the wave transmission through one-dimensional disordered media depend on the two-point correlations in the random scattering potential. Therefore, if the power spectrum $S(2k)$ is very small or vanishes within some interval of the wave number $k$, then the localization length $L_{loc}$ appears to be very large ($L_{loc}\gg L$) or even diverges. Evidently, the localization effects can be neglected in this case, and the structure, even of a large length, is fully transparent. This means that, in principle, by a proper choice of the disorder one can design the disordered structures with selective ({\it anomalous}) ballistic transport within a prescribed range of $k$.

Taking into account the form \eqref{SD-SM-V} of the
potential and its correlation properties \eqref{SD-Ksi}, from Eqs.~\eqref{1D-Lloc} one can readily derive the
following explicit formula for the localization length in the single-mode waveguide \cite{MY89},

\begin{equation}\label{SD-SM-Lloc}
L_{loc}^{-1}(k_1)=\frac{2\sigma^2}{\pi^2}
\left(\frac{\pi}{d}\right)^6\frac{W(2k_1)}{(2k_1)^2}.
\end{equation}
Since the potential \eqref{SD-SM-V} is entirely determined by the
rough surface profile $\xi(x)$, the localization length \eqref{SD-SM-Lloc} is specified by the roughness power spectrum $W(k_x)$. Therefore, by a proper fabrication of a random profile $\xi(x)$ with specific long-range correlations, one can arrange a desirable anomalous transport within a given window of $k=\omega/c$ inside the single-mode region \eqref{SM-k1}.

Before we start with a practical implementation of the
expression \eqref{SD-SM-Lloc} for the inverse localization length, it is worthwhile to clear up some points. First, one should stress that this result, as well as Eq.~\eqref{1D-Lloc} obtained by different methods, is an asymptotic one. This means that the higher terms are non-controlled; however, they can be neglected in the limit $\sigma^2\to0$. Second, the main assumptions used in the derivation of Eq.~\eqref{SD-SM-Lloc} are based on the validity of averaging over different realizations of disorder (see, for instance, Ref.~\cite{M99}). The condition for such an average is that two lengths, $L_{loc}$ and $L$, are much larger than two other characteristic lengths, the wavelength $k_1^{-1}$ and the correlation length $k_c^{-1}$ determining the maximal value of the power spectrum $W(2k_1)$. One should stress that for any finite values of $k_1$ and $k_c$, this condition can be always fulfilled due to an asymptotic character of Eq.~\eqref{SD-SM-Lloc} (or, the same, due to small value of $\sigma^2$).

\section{Design of random surface profile with predefined correlations: Convolution method}
\label{sec-ConvMethod}

From the above consideration it is seen that, in principle, by a proper choice of surface disorder one can artificially create the systems with selective transparency or reflectivity. Thus, the important practical problem arises of how to construct a corrugated surface profile from a predefined roughness power spectrum $W(k_x)$. This problem can be solved by employing a widely used \emph{convolution method} that was originally proposed in Ref.~\cite{R54}. The modern applications of this method for generation of random structures with specific correlations,
including long-range non-exponential correlations, can be found
in Refs.~\cite{S88,WOD95,CMHS95,BS95,RS99,CGK06,IKMU07} and in other papers cited in this chapter.

The method consists of the following steps. First, having a desirable form for the power spectrum $W(k_x)$, we derive the \emph{modulation function} $\beta(x)$ whose Fourier transform is $W^{1/2}(k_x)$,

\begin{equation}\label{beta-def}
\beta(x)=\int_{-\infty}^{\infty}\frac{dk_x}{2\pi}
\exp\left(ik_xx\right)\,W^{1/2}(k_x).
\end{equation}
Then the random surface profile $\xi(x)$ is generated as a
convolution of a white-noise $Z(x)$ with the modulation function $\beta(x)$,

\begin{equation}\label{xi-beta}
\xi(x)=\sigma\,\int_{-\infty}^\infty\,dx'\,Z(x-x')\,\beta(x').
\end{equation}
The delta-correlated random process $Z(x)$ is determined by the
standard relations,

\begin{equation}\label{Zcorr}
\langle Z(x)\rangle=0, \qquad\qquad
\langle Z(x)Z(x')\rangle=\delta(x-x'),
\end{equation}
and can be numerically created with the use of random-number generators. Here, $\delta(x)$ is the Dirac delta-function.

The expressions \eqref{xi-beta} and \eqref{Zcorr} give the solution of the inverse scattering problem of constructing random roughness from its power spectrum. Note that this method is valid in the case of a weak disorder only. That is why only the binary correlator is
involved in the construction of $\xi(x)$ while the higher
correlators do not contribute. Note also that the profile
obtained by the proposed method is not unique. Indeed, there is an
infinite number of realizations of delta-correlated noise $Z(x)$ that give rise to different profiles $\xi(x)$ having the same power spectrum $W(k_x)$.

The importance of this method is due to a possibility to obtain profiles resulting in a sharp transition between ballistic and localized transport, when changing the wave number $k$. In this case the corresponding power spectrum $W(k_x)$ abruptly vanishes at prescribed values of $k_c$. This means that the binary correlator ${\cal W}(x)$ has to be a slowly decaying function of the distance $|x|$. In other words, the corresponding corrugated surface profiles $\xi(x)$ should be of specific form revealing long-range correlations along the structure. Because of an abrupt character of transmission properties, the transition point $k_c$ can be regarded as an effective {\it transparency edge}.

As is pointed out above, a statistical treatment is meaningful if
the scale of decrease $k_c^{-1}$ of the correlator ${\cal
W}(x)$ is much less than both the sample length $L$ and
localization length $L_{loc}$. In this connection one should stress that long-range correlations we speak about, do not assume large values for the correlation length $k_c^{-1}$. Indeed, the simplest correlator, ${\cal W}(x)=\sin{k_cx}/k_cx$ has finite scale $k_c^{-1}$ of its decrease and $k_c^{-1}$ can be quite small, see examples below. On the other hand, an effective width of the transparency edge is determined by the product, $(L_{loc}k_c)^{-1}$, not by $k_c^{-1}$, and turns out to be very small (for details, see, e.g., Ref.~\cite{M99}). One can say that the sharpness of the transition is defined by the form of the pair correlator rather than by the value of its correlation length.

Note that the systems with very complicated scattering potentials are not exotic. For example, bulk random potentials have been constructed in the experiments \cite{KIK08,KIKS00,KIKSU02,KS98}, while rough surfaces with rectangular power spectrum have been fabricated in the study of the backscattering enhancement \cite{WOD95}.

\section{Gaussian correlations}

As was mentioned above, the general expression \eqref{SD-SM-Lloc} for the localization length $L_{loc}(k_1)$ indicates that all features of the wave transmission through the surface-disordered single-mode waveguide depend on two-point correlations in the surface profile. In order to demonstrate how to realize the properties of long-range correlated disorder, let us first
consider the surface roughness $\xi(x)$ with a widely used Gaussian correlator,

\begin{subequations}\label{GausCorPS}
\begin{eqnarray}
{\cal W}(x)&=&\exp\left(-k_c^2x^2\right),\\[6pt]
W(k_x)&=&\sqrt{\pi}\,k_c^{-1}\exp\left(-k_x^2/4k_c^2\right).
\end{eqnarray}
\end{subequations}
This correlator exponentially decreases on the scale of the correlation length $k_c^{-1}$.

Using the convolution method \eqref{beta-def} -- \eqref{Zcorr} one can obtain that the corrugated surface profile $\xi(x)$ with the correlation properties \eqref{GausCorPS} is described by the function,

\begin{equation}\label{xi-Gaus}
\xi(x)=\frac{\sigma\sqrt{2k_c}}{\pi^{1/4}}\, \int_{-\infty}^\infty\,dx'\,Z(x-x')\,\exp\left(-2k_c^2x'^2\right).
\end{equation}
Correspondingly, the inverse localization length \eqref{SD-SM-Lloc} takes the following explicit form,

\begin{equation}\label{SD-LlocGaus}
L_{loc}^{-1}(k_1)=\frac{2\sigma^2}{\pi\sqrt{\pi}k_c}
\left(\frac{\pi}{d}\right)^6\frac{\exp\left(-k_1^2/k_c^2\right)}{(2k_1)^2}.
\end{equation}
One can see that within the single-mode interval \eqref{SM-k1} the localization length increases exponentially with $k_1$ from zero at $k_1=0$ to a large value at $k_1=\pi\sqrt{3}/d$. Clearly, in the vicinity of $k_1=0$ the localization length is $L_{loc}(k_1)$ is much less than $L$ and the waveguide is non-transparent. Thus, the localization regime \eqref{1D-Tloc} occurs within the whole single-mode interval, provided by the condition $L_{loc}(\pi\sqrt{3}/d)\ll L$ at $k_1=\pi\sqrt{3}/d$.

On the contrary, when $L_{loc}(\pi\sqrt{3}/d)\gg L$, one can observe the crossover from the localized transport \eqref{1D-Tloc} to the ballistic one \eqref{1D-Tbal}. For the Gaussian correlations \eqref{GausCorPS}, both the crossing point where $L_{loc}(k_1)=L$, and the crossover width depend on the values of $k_c^{-1}$ and $L$. Because of a smooth character of the crossover, the crossing point cannot be regarded as the transparency edge. However, one can see that the longer the correlation length $k_c^{-1}$, the larger localization length $L_{loc}(k_1)$. Hence, the larger is the ballistic region and more narrow is the crossover. One can conclude that, in general, \emph{the correlations suppress the localization}.

The surface profile $\xi(x)$ with Gaussian correlations admits the uncorrelated roughness of the white-noise type. Indeed, since $\sigma^2k_c^{-1}=\mathrm{const}$, from Eqs.~\eqref{GausCorPS} for $k_c^{-1}\rightarrow 0$ one obtains the delta-like correlator and constant power spectrum,

\begin{subequations}\label{WNCorPS}
\begin{eqnarray}
{\cal W}_{wn}(x)&=&\sqrt{\pi}\,k_c^{-1}\delta(x),\\[6pt]
W_{wn}(k_x)&=&\sqrt{\pi}\,k_c^{-1}.
\end{eqnarray}
\end{subequations}

The convolution method \eqref{beta-def} -- \eqref{Zcorr} results in the following expression for the surface profile $\xi(x)$:

\begin{equation}\label{xi-WN}
\xi_{wn}(x)=\frac{\sigma\pi^{1/4}}{\sqrt{k_c}}\,Z(x).
\end{equation}
According to Eq.~\eqref{SD-SM-Lloc}, the localization length reads,

\begin{equation}\label{SD-LlocWN}
\frac{1}{L_{loc}^{wn}(k_1)}=\frac{\sigma^2k_c^{-1}}{2\pi\sqrt{\pi}k_1^2}\,
\left(\frac{\pi}{d}\right)^6.
\end{equation}
The expressions \eqref{WNCorPS} -- \eqref{SD-LlocWN} can be considered as the asymptotical ones of the respective equations \eqref{GausCorPS} -- \eqref{SD-LlocGaus} when $(k_1/k_c)^2\ll1$.

The comparison of Eqs.~\eqref{SD-LlocWN} and \eqref{SD-LlocGaus} leads to the conclusion: the best way to observe localized transport is to employ an uncorrelated disordered surface. Indeed, the condition $L_{loc}(k_1)\ll L$ is stronger for the Gaussian correlations than for the case of white-noise profiles. On the other hand, for Gaussian correlations with a small value of $k_c$, the ballistic regime \eqref{1D-Tbal} can be realized even for such lengths $L$ and wave numbers $k_1$ for which strong localization, \eqref{1D-Tloc}, takes place for delta-like correlations. Again, this fact confirms that the correlations suppress the localization.

\section{Two complementary examples of selective transparency}

As we discussed in Section~\ref{sec-ConvMethod}, the rough surfaces with prescribed two-point correlations can be constructed with the use of the convolution method. Below we demonstrate the construction of surface-disordered structures with selective transparency, by considering two examples of long-range correlations.

\textbf{(a)} Let us first consider the waveguide which is
non-transparent when the wave number $k_1$ is less than some value
$k_c$, and completely transparent for $k_1>k_c$. Such a
behavior can be observed if the transition point (transparency edge) $k_1=k_c$ is located inside the allowed single-mode interval
\eqref{SM-k1},

\begin{equation}\label{SM-kc}
0<k_cd/\pi<\sqrt{3}.
\end{equation}
For this case one can get the following expressions for the binary
correlator ${\cal W}(x)$ and power spectrum $W(k_x)$:

\begin{subequations}\label{SM-Wa}
\begin{eqnarray}
{\cal W}_{a}(x)&=&\frac{\sin(2k_cx)}{2k_cx},\\[6pt]
W_{a}(k_x)&=&\frac{\pi}{2k_c}\,\Theta(2k_c-|k_x|).
\end{eqnarray}
\end{subequations}
Here $\Theta(x)$ is the Heaviside unit-step function,
$\Theta(x<0)=0$ and $\Theta(x>0)=1$, and $k_c$ is the correlation parameter to be specified.

According to the recipe \eqref{beta-def} -- \eqref{Zcorr}, the surface profile with the above properties is given by the expression

\begin{equation}\label{SM-xia}
\xi_{a}(x)=\frac{\sigma}{\sqrt{2\pi k_c}}
\int_{-\infty}^{\infty}dx'Z(x-x')\frac{\sin(2k_cx')}{x'}.
\end{equation}
Correspondingly, the inverse localization
length has the {\it step-down} form,

\begin{equation}\label{SM-Lloca}
L_{loc}^{-1}(k_1)=\frac{\sigma^2}{\pi k_c}
\left(\frac{\pi}{d}\right)^6 \frac{\Theta(k_c-k_1)}{(2k_1)^2}.
\end{equation}
In line with this expression, as $k_1$ increases, the localization
length $L_{loc}(k_1)$ also smoothly increases and then diverges
at $k_1=k_c$. Thus, within the region $0<k_1<k_c$ the average
transmittance $\langle T(L/L_{loc})\rangle$ is expected to be
exponentially small \eqref{1D-Tloc} due to strong localization. The condition for strong localization to the left from $k_1=k_c$ reads,

\begin{equation}\label{SM-loca}
\frac{L}{L_{loc}(k_c-0)}=\frac{\sigma^2L}{4\pi k_c^3}\,
\left(\frac{\pi}{d}\right)^6\gg1.
\end{equation}
Otherwise, inside the interval $k_c<k_1<\pi\sqrt{3}/d$ a ballistic regime occurs with perfect transparency, $\langle T(L/L_{loc})\rangle=1$.

\textbf{(b)} The second example refers to a complementary situation when for $k_1<k_c$ the waveguide is perfectly transparent and for
$k_1>k_c$ is non-transparent. The corresponding expressions for
${\cal W}(x)$ and $W(k_x)$ are given by,

\begin{subequations}\label{SM-Wb}
\begin{eqnarray}
{\cal W}_b(x)&=&\pi\delta(2k_cx)-\frac{\sin(2k_cx)}{2k_cx},\\[6pt]
W_b(k_x)&=&\frac{\pi}{2k_c}\,\Theta(|k_x|-2k_c).
\end{eqnarray}
\end{subequations}
In this case the corrugated surface is described by a superposition of a white noise and the roughness of the first type,

\begin{equation}\label{SM-xib}
\xi_b(x)=\frac{\sigma}{\sqrt{2\pi k_c}}\left[\pi Z(x)-
\int_{-\infty}^{\infty}dx'Z(x-x')\frac{\sin(2k_cx')}{x'}\right].
\end{equation}
Correspondingly, the inverse localization length is expressed by
the \emph{step-up} function,

\begin{equation}\label{SM-Llocb}
L_{loc}^{-1}(k_1)=\frac{\sigma^2}{\pi
k_c}\left(\frac{\pi}{d}\right)^6 \frac{\Theta(k_1-k_c)}{(2k_1)^2}.
\end{equation}
As a consequence, in contrast with the first case, here the
surface-scattering localization length $L_{loc}(k_1)$ diverges
below the transparency edge $k_1=k_c$. At this point $L_{loc}(k_1)$
sharply falls down to a finite value $L_{loc}(k_c+0)$, and then
smoothly increases with further increase of $k_1$. In order to
observe the localization within the whole region
$k_c<k_1<\pi\sqrt{3}/d$, one should assume that strong localization is retained at upper point $k_1=\pi\sqrt{3}/d$ of the single-mode region \eqref{SM-k1},

\begin{equation}\label{SM-locb}
\frac{L}{L_{loc}(\pi\sqrt{3}/d)}=\frac{\sigma^2L}{12\pi k_c}\,
\left(\frac{\pi}{d}\right)^4\gg1.
\end{equation}
Therefore, in this example the ballistic transport is
abruptly replaced by strong localization at the transparency edge,
$k_1=k_c$.

One should stress that the surface profiles \eqref{SM-xia} and \eqref{SM-xib} with respective binary correlators and power spectra \eqref{SM-Wa} and \eqref{SM-Wb} are substantially different from the delta-correlated white noise \eqref{xi-WN}, \eqref{WNCorPS}, and from random processes \eqref{xi-Gaus} with exponentially decaying Gaussian correlations \eqref{GausCorPS}. Specifically, here the profiles are characterized by the variation scale $(2k_c)^{-1}$ and have the long power-decaying tales in the expressions for their two-point correlator. Such tails are originated from the stepwise discontinuity at the points $k_x=\pm2k_c$. The location of the transparency edge is defined by these discontinuity points and do not depend on other parameters, in contrast to the case of Gaussian or delta-like correlations.

Now we demonstrate the above predictions by a direct numerical
simulation. For this, the inverse localization length
$L_{loc}^{-1}$ was computed with the use of the Hamiltonian map approach developed in Refs.~\cite{IK99,IKU01}. First, the continuous scattering potential \eqref{SD-SM-V} was approximated by the sum of delta kicks with the spacing $\delta$ chosen much smaller than any physical length scale in the model. After, a discrete analogue of the 1D wave equation \eqref{1DSchr-eq} was analyzed numerically, with the lengthwise wave number $k_1$ in place of $k$ and with the surface scattering potential $V(x)$ given by Eq.~\eqref{SD-SM-V}. In this way the wave equation was expressed in the form of a two-dimensional Hamiltonian map describing the dynamics of a classical linear oscillator under the parametric noise determined by $V(x)$. As a result, the analysis of the localization length was reduced to the computation of the Lyapunov exponent $L_{loc}^{-1}$ associated with this map (see details in Refs.~\cite{IK99,IKU01}).

\begin{figure}[h]
\begin{center}
\hspace{-0.5cm}
\subfigure{\includegraphics[width=2.45in,height=1.9in,angle=0]{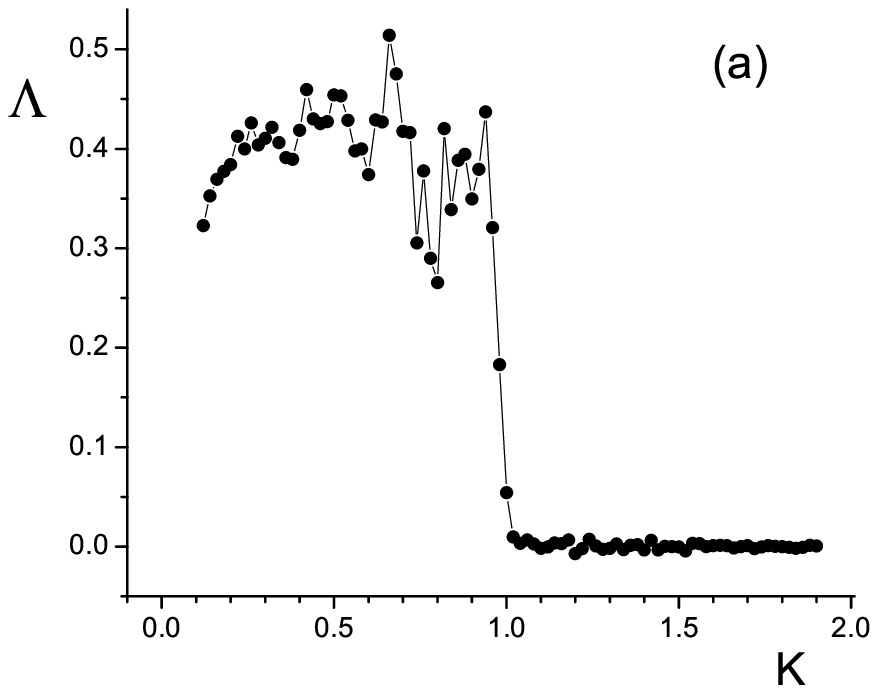}}
\subfigure{\includegraphics[width=2.45in,height=1.9in,angle=0]{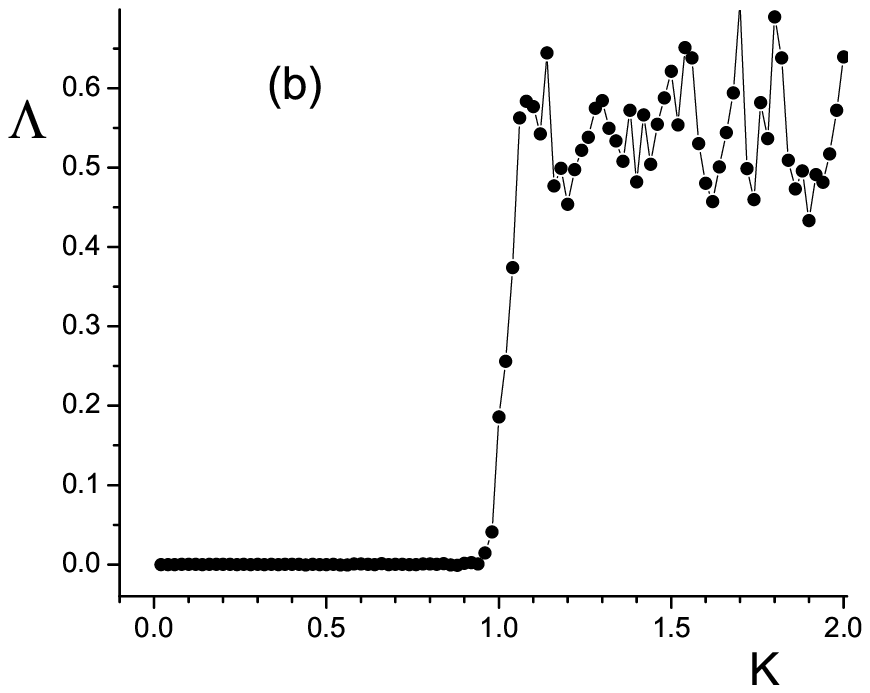}}
\vspace{-0.5cm}
\caption[Selective dependence of the rescaled Lyapunov exponent on wave number for two realizations of random surface with specific long-range correlations.]{Selective dependence of
the rescaled Lyapunov exponent on wave number for two
realizations of random surface with specific long-range
correlations. To show the main effect of correlations, the
complementary dependence of the Lyapunov exponent on $K$ is
shown: (a) eigenstates are localized for $K<1$ and
delocalized for $K>1$; (b) complete delocalization for
$K<1$ alternates by strong localization for $K>1$ (after \cite{IM05}).}
\label{Fig2ab-SMW}
\end{center}
\end{figure}

Numerical data reported in Figure~\ref{Fig2ab-SMW}, represent the
dependence of the dimensionless Lyapunov exponent
$\Lambda=c_0L_{loc}^{-1}$ on the normalized wave number
$K=k_1/k_c$ in the range $0<K<2$ corresponding to the single-mode
interval. The normalization coefficient $c_0$ was set to have
$\Lambda=K^{-2}$ for the delta-correlated potential. Two surface
profiles $\xi(x)$ were generated according to discrete versions of
Eqs.~\eqref{SM-xia} and \eqref{SM-xib} determining
complementary stepwise dependencies of the localization length
$L_{loc}(k_1)$ in accordance with Eqs.~\eqref{SM-Lloca} and
\eqref{SM-Llocb}.

One can clearly see a non-trivial dependence of $\Lambda$ on the
wave vector $K$, which is due to specific long-range correlations
in $\xi(x)$. The data display sharp dependencies of $\Lambda$ on
$K$ when crossing the point $K=1$. Thus, by taking the size $L$ of
the scattering region according to the requirements \eqref{SM-loca}) or \eqref{SM-locb}, one can arrange
anomalous transport in the single-mode guiding structure as is
predicted by the analytical theory.

\section{Random narrow-band reflector}

Let us now consider one more example of the binary correlator giving rise to the power spectrum of the following rectangular form,

\begin{subequations}\label{WOD-W}
\begin{eqnarray}
{\cal W}(x)&=&\frac{\sin(2k_+x)-\sin(2k_-x)}{2(k_+-k_-)x},\\[6pt]
W(k_x)&=&\frac{\pi}{2(k_+-k_-)}\left[\Theta(2k_+-|k_x|)-
\Theta(2k_--|k_x|)\right],\\[6pt]
&&0<k_-<k_+<\pi\sqrt{3}/d.\nonumber
\end{eqnarray}
\end{subequations}
In these relations the factor $1/2(k_+-k_-)$ stands to provide the normalization requirement \eqref{W-norm}, or the same, ${\cal W}(0)=1$. Such a power spectrum has been employed to create specific rough surfaces in the experimental study of an enhanced backscattering \cite{WOD95}. This spectrum was also used in the
theoretical analysis of light scattering from amplifying media, as well as in the study of localization of plasmon
polaritons on random surfaces \cite{SLM01}.

In accordance with the convolution method \eqref{beta-def} -- \eqref{Zcorr}, the random surface profile $\xi(x)$ with the above correlation properties can be obtain from the following expression:

\begin{equation}\label{xi-WOD}
\xi(x)=\frac{\sigma}{\sqrt{2\pi}}\,\int_{-\infty}^\infty\,dx'\,Z(x-x')\,
\frac{\sin(2k_+x')-\sin(2k_-x')}{(k_+-k_-)^{1/2}x'}.
\end{equation}
The peculiarity of such surfaces is that they have two characteristic scales, $(2k_+)^{-1}$ and $(2k_-)^{-1}$. Consequently, the binary correlator and its power spectrum \eqref{WOD-W} are specified by two correlation parameters.

From Eqs.~\eqref{SD-SM-Lloc} and \eqref{WOD-W} one can find
the inverse localization length,

\begin{equation}\label{Lloc-WOD}
L_{loc}^{-1}(k_1)=\frac{\sigma^2}{\pi(k_+-k_-)}
\left(\frac{\pi}{d}\right)^6\frac{\Theta(k_+-k_1)\Theta(k_1-k_-)}{(2k_1)^2}.
\end{equation}
As one can see, there are two transparency edges at the points $k_1=k_-$ and $k_1=k_+$. The localization length $L_{loc}(k_1)$ diverges below the first point, $k_1=k_-$, and above the second one, $k_1=k_+$. Between these points, for $k_-<k_1<k_+$, the localization length \eqref{Lloc-WOD} has finite value and smoothly increases with an increase of wave number $k_1$.

Let us now choose the parameters for which the regime of a strong
localization occurs at the upper transition point $k_1=k_+$ where $L_{loc}(k_1)$ gets its maximal value. This
automatically provides strong localization within the whole
interval $k_-<k_1<k_+$. The condition to have this situation reads

\begin{equation}\label{LocTr-WOD}
\frac{L}{L_{loc}(k_+-0)}=\frac{\sigma^2L}{4\pi k_+^2(k_+-k_-)} \left(\frac{\pi}{d}\right)^6\gg1.
\end{equation}
As a result, there are two regions of perfect transparency for the waveguides of finite length $L$ with the specified above surface profile. Between these regions the average transmittance $\langle T\rangle$ is exponentially small according to the expression \eqref{1D-Tloc}. Due to this fact, the system exhibits the localized transport within the interval $k_-<k_1<k_+$ and the
ballistic regime with $\langle T\rangle=1$ outside this interval.
In the experiment one can observe that with an increase of the
wave number $k_1$, the perfect transparency below $k_1=k_-$ abruptly alternates with a complete reflection, and recovers at $k_1=k_+$.
From equations \eqref{Lloc-WOD} and \eqref{LocTr-WOD} one can see
that the smaller value $k_+-k_-$ of the reflecting region the
smaller the surface-scattering localization length $L_{loc}(k_1)$ and, consequently, the stronger is the localization within this region. This remarkable fact may find important applications in creating a new class of random narrow-band filters or reflectors.

\section{Multi-mode waveguide}

Now we examine the correlated surface scattering in \emph{multi-mode} waveguides, i.e. in waveguides with a large number $N_d>1$ of conducting channels, see expression \eqref{Nd}. According to Landauer's concept \cite{L92}, the \emph{total average transmittance} $\langle T\rangle$ of any disordered quasi-one-dimensional guiding structure can be expressed as a sum of \emph{partial average transmittances} $\langle T_n\rangle$ that describe the transport for every $n$-th propagating normal mode,

\begin{equation}\label{SD-MM-T}
\langle T\rangle=\sum_{n=1}^{N_d}\langle T_n\rangle.
\end{equation}
When all conducting channels are open, i.e. all $T_n=1$, the total transmittance \eqref{SD-MM-T} attains its maximal value equal to the total number $N_d$ of propagating modes. Therefore, our definition of the transmittance differs from the canonical one in which the maximal value of total transmittance is equal to one. Nevertheless, we shall use the definition \eqref{SD-MM-T} in order to clearly discriminate the intervals of the mode parameter $kd/\pi$ with different numbers of the conducting channels and more clearly display the role of correlated surface disorder. In order to pass to the definition used in wave theories, one should simply divide our expression \eqref{SD-MM-T} over $N_d$.

From general theory of quasi-one-dimensional scattering systems it follows that the transmission properties of any $n$th conducting channel ($1\leqslant n\leqslant N_d$) are determined by \emph{two} attenuation lengths, the \emph{forward scattering length} $L_n^{(f)}$ and \emph{backscattering length} $L_n^{(b)}$. For multi-mode quasi-one-dimensional waveguides with surface disorder the inverse scattering lengths are given by,

\begin{equation}\label{SD-MM-Lnf}
\frac{1}{L_n^{(f)}}=\sigma^2\frac{(\pi n/d)^2}{2k_nd}
\sum_{n'=1}^{N_d}\frac{(\pi n'/d)^2}{k_{n'}d}W(k_n-k_{n'}),
\end{equation}

\begin{equation}\label{SD-MM-Lnb}
\frac{1}{L_n^{(b)}}=\sigma^2\frac{(\pi n/d)^2}{2k_nd}
\sum_{n'=1}^{N_d}\frac{(\pi n'/d)^2}{k_{n'}d}W(k_n+k_{n'}).
\end{equation}
Here the lengthwise wave number $k_n$ is defined by Eq.~\eqref{kn}. The results \eqref{SD-MM-Lnf} and \eqref{SD-MM-Lnb}
can be obtained for the boundary-value problem \eqref{2D-BVP} by
the diagrammatic Green's function method \cite{BFb79}, as well as
by the technique developed in Ref.~\cite{MM84}. Also,
these expressions can be derived by using the invariant imbedding
method extended to quasi-one-dimensional structures \cite{SFYM98}. Note that in a single-mode waveguide with $N_d=1$ the sum over $n'$ contains only one term with $n'=n=1$. Therefore, in this case the backscattering length $L_1^{(b)}$ is exactly equal to the single-mode localization length \eqref{SD-SM-Lloc},
$L_1^{(b)}=L_{loc}(k_1)$, see Eq.~\eqref{<1D-Ts>}.

The expressions \eqref{SD-MM-Lnf} and \eqref{SD-MM-Lnb} manifest
that, in general, both attenuation lengths are contributed by
scattering of a given $n$-th propagating mode into all other
modes. This is the case when, for example, a rough surface profile
is either delta-correlated random process with
constant power spectrum, $W(k_x)=const$, or has fast decreasing
binary correlator ${\cal W}(x)$ and, correspondingly, slowly decreasing roughness power spectrum $W(k_x)$.

Another peculiarity is that the expressions \eqref{SD-MM-Lnf} and
\eqref{SD-MM-Lnb} display rather strong dependence on the mode
index $n$. Namely, the larger the number $n$ the smaller the
corresponding mode scattering lengths and as a consequence, the
stronger is the scattering of this mode into the others. This strong dependence is due to squared transverse wave number $k_z=\pi n/d$ in the numerator and lengthwise wave number $k_n$ in the denominator of Eqs.~\eqref{SD-MM-Lnf} and \eqref{SD-MM-Lnb}. Evidently, with an increase of the mode index $n$ the value of $k_n$ decreases. An additional dependence appears because of the roughness power spectrum $W(k_n\mp k_{n'})$. Since the binary correlator ${\cal W}(x)$ of random surfaces is a decreasing function of $|x|$, its Fourier transform $W(k_n\mp k_{n'})$ increases with $n$ (note that it is constant for the delta-correlated roughness only). Therefore, all the factors contribute in the same direction for the dependence of $L_n^{(f)}$ and $L_n^{(b)}$ on the mode index $n$. As a result, we arrive at  the following hierarchy of mode scattering lengths:

\begin{equation}\label{SD-MM-Ln-hierarchy}
L_{N_d}^{(f,b)}<L_{N_d-1}^{(f,b)}<...<L_2^{(f,b)}<L_1^{(f,b)}.
\end{equation}
The smallest mode attenuation lengths $L_{N_d}^{(f)}$ and
$L_{N_d}^{(b)}$ belong to the highest channel with the
mode index $n=N_d$, while the largest scattering lengths
$L_1^{(f)}$ and $L_1^{(b)}$ correspond to the lowest
channel with $n=1$. Note that similar hierarchy
was also found in Refs.~\cite{IM04a,IM04b,IM05} in the model of quasi-one-dimensional systems with a stratified disorder.

As is known, the quasi-one-dimensional systems with isotropic volume disorder reveal three typical transport regimes, the regimes of a
\emph{ballistic}, \emph{diffusive} (metallic), and \emph{localized}
transport. In contrast to this conventional picture, in Refs.~\cite{SFYM98} it was shown that in the case of surface disorder a very important phenomenon of the \emph{coexistence} of ballistic, diffusive, and localized transport emerges. This happens due to hierarchy \eqref{SD-MM-Ln-hierarchy} of scattering lengths, and even in the absence of correlations in $\xi(x)$,  Specifically, while the lowest modes can be in the ballistic regime, the intermediate and highest modes can exhibit the diffusive and localized behavior, respectively. This effect seems to be generic for the transport through the waveguides with random surfaces.

One can see now that unlike the single-mode case, the concept of
one-parameter scaling is no more valid for the transport in
multi-mode surface-disordered systems. There are two points that should be stressed in this respect. On the one hand, the average partial transmittances $\langle T_n\rangle$ entering the equation
\eqref{SD-MM-T}, are very different for different conducting
channels. On the other hand, and what is even more important, all
propagating modes turn out to be mixed due to the inter-mode
transitions. Therefore, the transmittance $\langle T_n\rangle$ of
a given $n$-th mode depends on the scattering into all modes, and the total average transmittance \eqref{SD-MM-T} is determined by the whole set of attenuation lengths \eqref{SD-MM-Lnf} and \eqref{SD-MM-Lnb} with $1\leqslant n\leqslant N_d$.

Summarizing our brief discussion, it becomes clear that in quasi-one-dimensional guiding structures with delta-correlated or Gaussian
correlations in surface disorder, the crossover from the ballistic to localized transport is realized through the successive localization of highest propagating modes. Otherwise, if we start from the localized regime, the crossover to the ballistic
transport is realized via the successive opening (delocalization)
of lowest conducting channels.

From this analysis one can conclude that for multi-mode structures
with surface disorder the role of specific long-range correlations
is much more sophisticated in comparison with single-mode waveguides. First, such
correlations should result in the suppression of the interaction
between different propagating modes. This non-trivial fact turns
out to be crucial for the reduction of a system of mixed channels
with quasi-one-dimensional transport to the subset of independent waveguide
modes with a purely 1D transport. Second, the same correlations
can provide a complete transparency of each independent channel,
similar to what happens in strictly 1D geometry.

To demonstrate these effects, let us take a random surface profile
$\xi(x)$ of the form,

\begin{equation}\label{SD-MM-xi}
\xi(x)=\frac{\sigma}{\sqrt{\pi k_c}}\,\int_{-\infty}^\infty\,dx'\,
Z(x-x')\,\frac{\sin(k_cx')}{x'}\,,
\end{equation}
with the slowly decaying (on average) binary correlator. It
results in the ``window function'' for the roughness power
spectrum,

\begin{subequations}\label{SD-MM-W}
\begin{eqnarray}
{\cal W}(x)&=&\frac{\sin(k_cx)}{k_cx}\,,\\[6pt]
W(k_x)&=&\frac{\pi}{k_c}\,\Theta(k_c-|k_x|),\qquad\qquad k_c>0.
\end{eqnarray}
\end{subequations}

From equations \eqref{SD-MM-Lnf} and \eqref{SD-MM-Lnb} one can see
that in the case of long-range correlations in a disordered
surface \eqref{SD-MM-xi}, the number of modes into which a given
$n$th mode is scattered, i.e. the actual number of summands in
the equations \eqref{SD-MM-Lnf} and \eqref{SD-MM-Lnb}, is entirely
determined by the width $k_c$ of the rectangular power spectrum
\eqref{SD-MM-W}. It is clear that if the distance
$|k_n-k_{n\pm1}|$ between neighboring quantum values of $k_n$ is
larger than the correlation width $k_c$,

\begin{equation}\label{SD-MM-cond}
|k_n-k_{n\pm1}|>k_c\,,
\end{equation}
then all inter-mode transitions (between different propagating modes) are forbidden. As a consequence, the sum over $n'$ in the expression \eqref{SD-MM-Lnf} for the inverse forward scattering
length contains only diagonal term with $n'=n$ describing a
direct intra-mode scattering \emph{inside} the channels. Moreover,
each term in the sum of Eq.~\eqref{SD-MM-Lnb} for the inverse
backscattering length is equal to zero. As a result, the following
interesting phenomena arise.

\textbf{(i)} All \emph{high} propagating modes with indices $n$ that satisfy the condition \eqref{SD-MM-cond}, turn out to be
independent of the others in spite of the interaction with rough
surface. Therefore, they form a subset of 1D non-interacting
conducting channels with finite length of forward scattering
$L_n^{(f)}$ and infinite backscattering length $L_n^{(b)}$,

\begin{equation}\label{SD-MM-Lnfb-CD}
\frac{1}{L_n^{(f)}}=\frac{\pi\sigma^2}{k_c}\,\frac{(\pi
n/d)^4}{(k_nd)^2}\,, \qquad\qquad\frac{1}{L_n^{(b)}}=0.
\end{equation}

\textbf{(ii)} As is well known from the standard theory of one-dimensional
localization (see, e.g., Refs.~\cite{B73,LGPb88,MKb04,MY89,MT98,M99}), the transport through any 1D disordered system is determined only by
the backscattering length $L_n^{(b)}$ that in our consideration equals the localization length, and does not depend on the
forward-scattering length $L_n^{(f)}$. Since the former diverges
for every independent channel in line with the expression
\eqref{SD-MM-Lnfb-CD}, all of them are completely transparent because they exhibit the ballistic transport with the partial average transmittance $\langle T_n\rangle=1$. This means that according to the Landauer's formula \eqref{SD-MM-T}, the transmittance of the subset of such independent ballistic modes is simply equal to their total number.

\textbf{(iii)} As for \emph{low} propagating modes with the indices $n$ contradicting to the condition \eqref{SD-MM-cond}, they remain to be mixed by surface scattering because the roughness power
spectrum \eqref{SD-MM-W} is non-zero for them, $W(k_n-k_{n'})=\pi/k_c$. These \emph{mixed modes} have finite
forward and backscattering lengths and, consequently, stay in the diffusive or localized transport regime for large enough
waveguide length $L$. As a result, they are non-transparent and do not contribute to the total transmittance $\langle T\rangle$. Therefore, the latter is equal to the number of independent ballistic modes.

Note that the distance $|k_n-k_{n\pm1}|$ between neighboring
wave numbers $k_n$ and $k_{n\pm1}$ increases as the mode
index $n$ increases. Therefore, the inequality \eqref{SD-MM-cond}
restricts the mode index $n$ from below. That is why, in contrast
with the conventional situation associated with the hierarchy of
mode scattering lengths \eqref{SD-MM-Ln-hierarchy}, the low modes
are mixed and non-transparent, while high propagating modes are
independent and ballistic. Because of a sharp behavior of the
roughness power spectrum \eqref{SD-MM-W}, the transition from
mixed to independent modes is also sharp.

More analytical results can be obtained for waveguides with large
number of conducting channels $N_d$, if the quantum numbers $n$ of independent ballistic modes are also large:

\begin{equation}\label{Nd-large}
N_d=[kd/\pi]\approx kd/\pi\geqslant n\gg1.
\end{equation}
In this case the inequality \eqref{SD-MM-cond} is reduced to the requirement $|\partial k_n/\partial n|>k_c$ which can be rewritten in the following explicit form,

\begin{equation}\label{SD-MM-Nmix}
n>N_{mix}=\left[\frac{(kd/\pi)}{\sqrt{1+(k_cd/\pi)^{-2}}}\right].
\end{equation}
We recall that square brackets stand for the integer part of the
inner expression.

The condition \eqref{SD-MM-Nmix} determines the total number
$N_{mix}$ of mixed non-transparent modes, the total number
$N_{bal}=N_d-N_{mix}$ of independent ballistic modes, and the
critical value of the mode index $n$ that divides these two
groups. All propagating modes with $n>N_{mix}$ are independent and
fully transparent, otherwise, they are mixed for $n\leqslant N_{mix}$ and characterized by finite scattering lengths $L_n^{(f)}$ and $L_n^{(b)}$. Therefore, the total average transmittance
\eqref{SD-MM-T} of the multi-mode structure is,

\begin{equation}\label{SD-MM-Tcd}
\langle T\rangle=[kd/\pi]-\left[kd/\pi\alpha_c\right],\qquad
\alpha_c=\sqrt{1+(k_cd/\pi)^{-2}}\,.
\end{equation}

The numbers $N_{mix}$ and $N_{bal}$ of mixed non-transparent and
independent ballistic modes are governed by two parameters, the
mode parameter $kd/\pi$ and the dimensionless correlation
parameter $k_cd/\pi$. In the case of ``weak" correlations when
$k_cd/\pi\gg1$, the number of mixed modes $N_{mix}$ is of the
order of $N_d$,

\begin{equation}\label{SD-MM-Nmix-wc}
N_{mix}\approx\left[\left(\frac{kd}{\pi}\right)-
\frac{1}{2}\left(\frac{kd}{\pi}\right)
\left(\frac{k_cd}{\pi}\right)^{-2}\right]\quad\mbox{for}\quad
k_cd/\pi\gg1.
\end{equation}
Consequently, in this case the number of ballistic modes $N_{bal}$
is small, or there are no such modes at all. If the parameter
$k_cd/\pi$ tends to infinity, $k_cd/\pi\to\infty$, the rough
surface profile becomes white-noise-like and, naturally,
$N_{mix}\to N_d$.

The most appropriate case is when a random surface profile is
strongly correlated so that the correlation parameter is small,
$k_cd/\pi\ll1$. Then the number of mixed non-transparent modes
$N_{mix}$ is much less than the total number of propagating modes
$N_d$,

\begin{equation}\label{SD-MM-Nmix-sc}
N_{mix}\approx\left[\left(\frac{kd}{\pi}\right)
\left(\frac{k_cd}{\pi}\right)\right]\ll N_d\quad\mbox{for}\quad
k_cd/\pi\ll1.
\end{equation}
Therefore, the number of independent modes $N_{bal}$ is large.
When the correlation parameter $k_cd/\pi$ decreases and becomes
anomalously small, $k_cd/\pi<(kd/\pi)^{-1}\ll1$, the number of
mixed modes $N_{mix}$ vanishes and \emph{all modes} become
independent and perfectly transparent. Evidently, if the
correlation parameter $k_cd/\pi$ vanishes, $k_cd/\pi\to0$, the
roughness power spectrum \eqref{SD-MM-W}, becomes
delta-function-like and, as a consequence, $N_{mix}=0$. In this
case the correlated disorder results in a perfect transmission of
waves.

\begin{figure}[h]
\begin{center}
\includegraphics[width=4.2in,height=2.4in,angle=0]{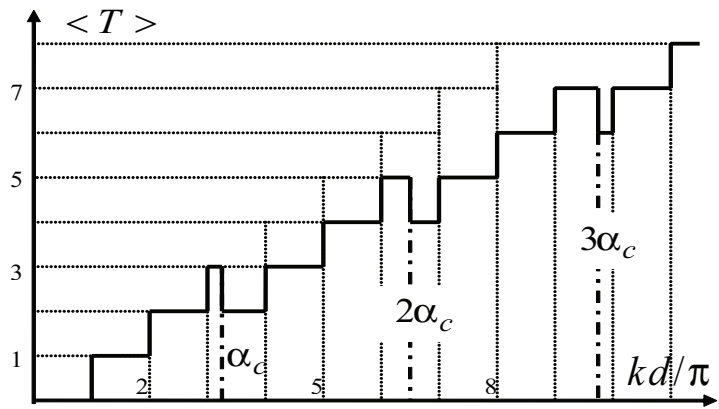}
\caption[Step-wise transmittance of surface-disordered multi-mode waveguide versus the mode parameter $kd/\pi$.]{Step-wise transmittance \eqref{SD-MM-Tcd} of surface-disordered guiding system versus the mode parameter $kd/\pi$. In the figure, the value of the normalized correlation parameter $k_cd/\pi=0.32$ (after \cite{IM05}).}
\label{Fig3-MMW}
\end{center}
\end{figure}

Finally, let us briefly discuss the expression \eqref{SD-MM-Tcd}
for the total average transmittance. In Figure~\ref{Fig3-MMW} an unusual \emph{non-monotonic stepwise dependence} of $\langle T\rangle$ on the mode parameter $kd/\pi$ is shown, that is governed by the width $k_c$ of the rectangular power spectrum \eqref{SD-MM-W}.

Let us discuss this Figure. Within the region where $kd/\pi<\alpha_c$, the inter-mode transitions caused by specific surface correlations, are forbidden for all conducting channels.
Therefore, all propagating modes are independent and ballistic,
and the second term (the number of mixed modes) in the expression
\eqref{SD-MM-Tcd} for the total transmittance is equal to zero.
Here the transmittance exhibits a ballistic \emph{stepwise increase} with an increase of the parameter $kd/\pi$. Each step ``up" arises
for an integer value of the mode parameter $kd/\pi$, when a new
conducting channel emerges. Such a stepwise increase of the total transmittance is similar to that known to occur in quasi-one-dimensional ballistic \emph{non-disordered} structures (see, e.g., \cite{Wo88}).

Otherwise, when $kd/\pi\geqslant\alpha_c$, in addition to the standard steps ``up" originated from the first term in equation
\eqref{SD-MM-Tcd}, there are also the steps ``down" associated with
the second term. These steps ``down" are provided by the correlated
surface scattering and arise when a successive low mode abruptly becomes mixed and non-transparent. In other words, the positions of the $n$th step ``down" are at the transparency edge point $kd/\pi=n\alpha_c$ where the $n$th conducting channels closes. Specifically, the first step ``down" occurs at the \emph{total transparency edge} $kd/\pi=\alpha_c$, where the first mode is closed. This transparency edge separates the region of complete transparency from that where lower modes are mixed and non-transparent. The second step ``down" is due to the \emph{particular transparency edge} $kd/\pi=2\alpha_c$ of the second mode, etc. Since the values of the ratio $kd/\pi\alpha_c$ are determined by the correlation parameter $k_c$, the positions of steps ``down", in general, do not coincide with those of steps ``up". The situation may also occur when the steps ``up" and ``down" cancel each other within some interval of the mode parameter $kd/\pi$. The interplay between steps ``up" and ``down" results in a new kind of \emph{stepwise non-monotonic dependence} of the total quasi-one-dimensional transmittance. The experimental observation of the discussed non-conventional dependence seems to be highly interesting.

F.M.I acknowledges the support by CONACyT grant No. 80715.


\end{document}